\RequirePackage[T1]{fontenc}
\documentclass[conference,letterpaper]{IEEEtran}
\usepackage{amsmath,amssymb}
\usepackage{graphicx}
\usepackage{xcolor}
\usepackage{booktabs,array,tabularx}
\usepackage{needspace}
\usepackage{cite}
\usepackage[hidelinks]{hyperref}
\usepackage{url}
\makeatletter
\newenvironment{inlinetable}{\par\noindent\begin{minipage}{\columnwidth}\def\@captype{table}}{\end{minipage}\par\vspace{8pt}}
\makeatother

\title{TernScope: Model--Hardware Co-Design\\for Efficient Ternary Endoscopic Segmentation}
\author{%
\IEEEauthorblockN{\normalsize Zhuoyu Wu\textsuperscript{1,*}, Wenhui Ou\textsuperscript{2}, Qiawei Zheng\textsuperscript{2}, Xuhong Qiu\textsuperscript{4}, Wenqi Fang\textsuperscript{3}, Raphael C.-W. Phan\textsuperscript{1}}
\IEEEauthorblockA{\footnotesize
\textsuperscript{1}CyPhi AI Research Lab, School of Information Technology, Monash University Malaysia, Malaysia\\
\textsuperscript{2}Department of Electronic and Computer Engineering, The Hong Kong University of Science and Technology, Hong Kong\\
\textsuperscript{3}Shenzhen Institutes of Advanced Technology, Chinese Academy of Sciences, Shenzhen, China\\
\textsuperscript{4}School of Computer Science and Engineering, Southeast University, Nanjing, China\\
\textsuperscript{*}Corresponding author: Zhuoyu Wu (\href{mailto:zhuoyu.wu@monash.edu}{zhuoyu.wu@monash.edu})\\
Code and experiments: \href{https://github.com/ReaganWu/TernScope}{\textcolor{orange!80!black}{\nolinkurl{https://github.com/ReaganWu/TernScope}}}}}
\newif\ifqualafterarchitecture
\qualafterarchitecturefalse
\begin{document}
\raggedbottom
\clubpenalty=10000
\widowpenalty=10000
\maketitle
\begin{abstract}
Edge medical segmentation must reconcile small model size, high accuracy, and fast inference. TernScope addresses this tradeoff through model--hardware co-design, learning a compact segmentation network whose low-bit operators also simplify hardware execution. Split-channel large-kernel blocks preserve spatial context with limited computation, while branch fusion removes redundant inference paths. Ternary Progressive Quantization-aware Training (TPQ) then adapts the fused network from deep to shallow layers, through the decoder, and finally to the input and prediction layers. All 25 convolutions use ternary weights and 8-bit activations, enabling packed storage and multiplier-free computation. The resulting model uses 0.2483\,M parameters and 0.0949\,GMACs while retaining competitive segmentation accuracy on in-domain and external datasets. Under the same 45 nm technology, TernScope reduces compute-array area by 88.46\% compared with the original Gemmini, a systolic-array accelerator with 8-bit $\times$ 8-bit multipliers. These results demonstrate a path from accurate lightweight medical segmentation to compact low-bit hardware.
\end{abstract}

\section{Introduction}
Medical image segmentation at the edge faces a \emph{model size--accuracy--inference speed trilemma}: a deployable model must fit limited memory, delineate lesions accurately, and deliver predictions promptly~\cite{valanarasu2022unext,karmakar2022mobilepolypnet,wu2027depthpolyp}. Reducing network width or spatial processing can save resources, but also restrict the contextual information available for segmentation~\cite{tang2024cmunext,chen2024tinyunet}. The goal is therefore to retain strong segmentation capacity within a small, fast model. For endoscopic applications, this requires treating predictive quality and deployment cost as joint design objectives.

\begin{figure}[t]
\centering
\includegraphics[width=0.84\columnwidth]{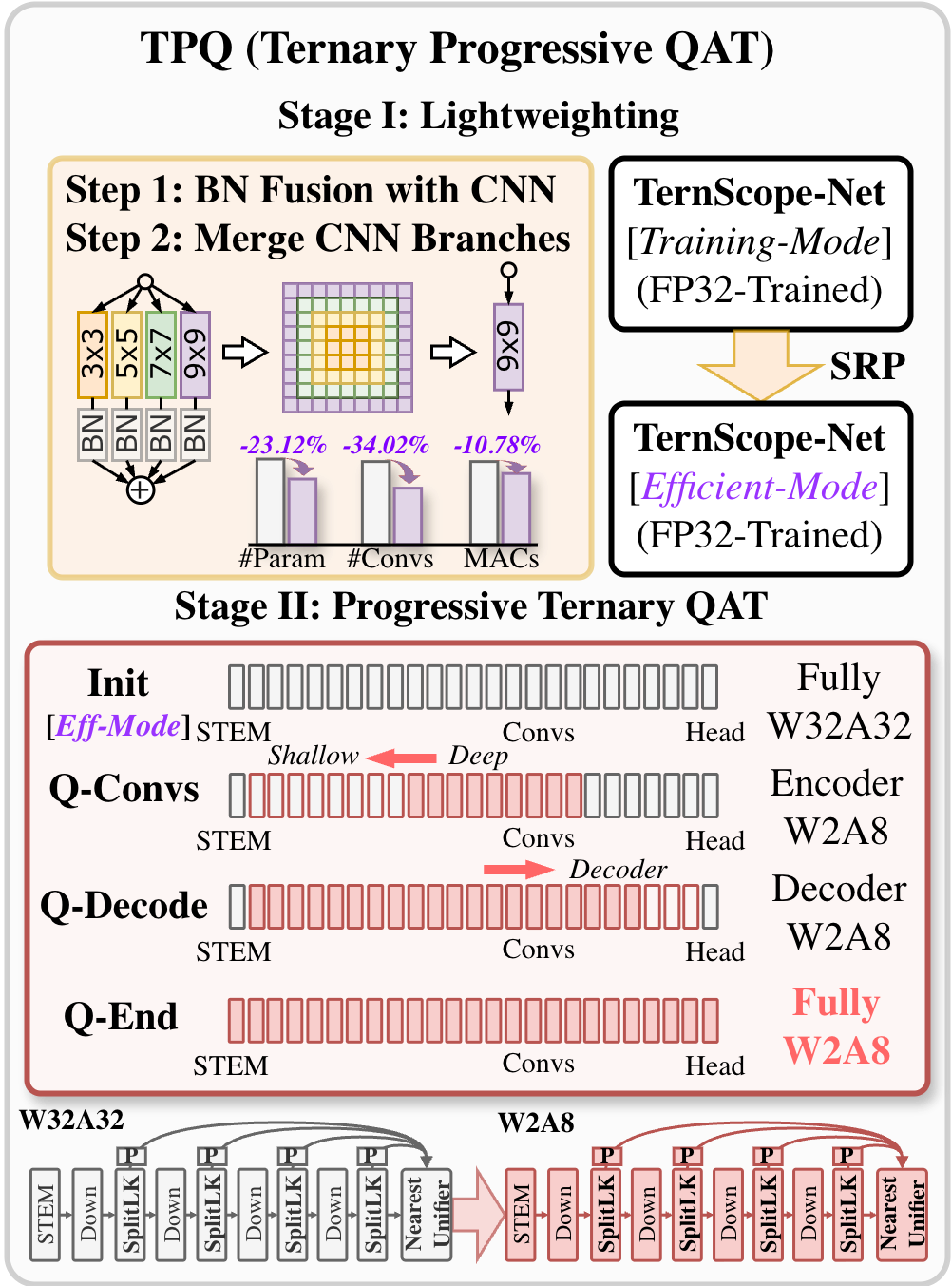}
\caption{TernScope training-to-deployment pipeline. Branch fusion removes redundant inference paths; TPQ adapts the fused graph to W2A8, including the stem and head.}
\label{fig:qat}
\end{figure}
\begin{figure*}[t]
\centering
\includegraphics[width=0.96\textwidth]{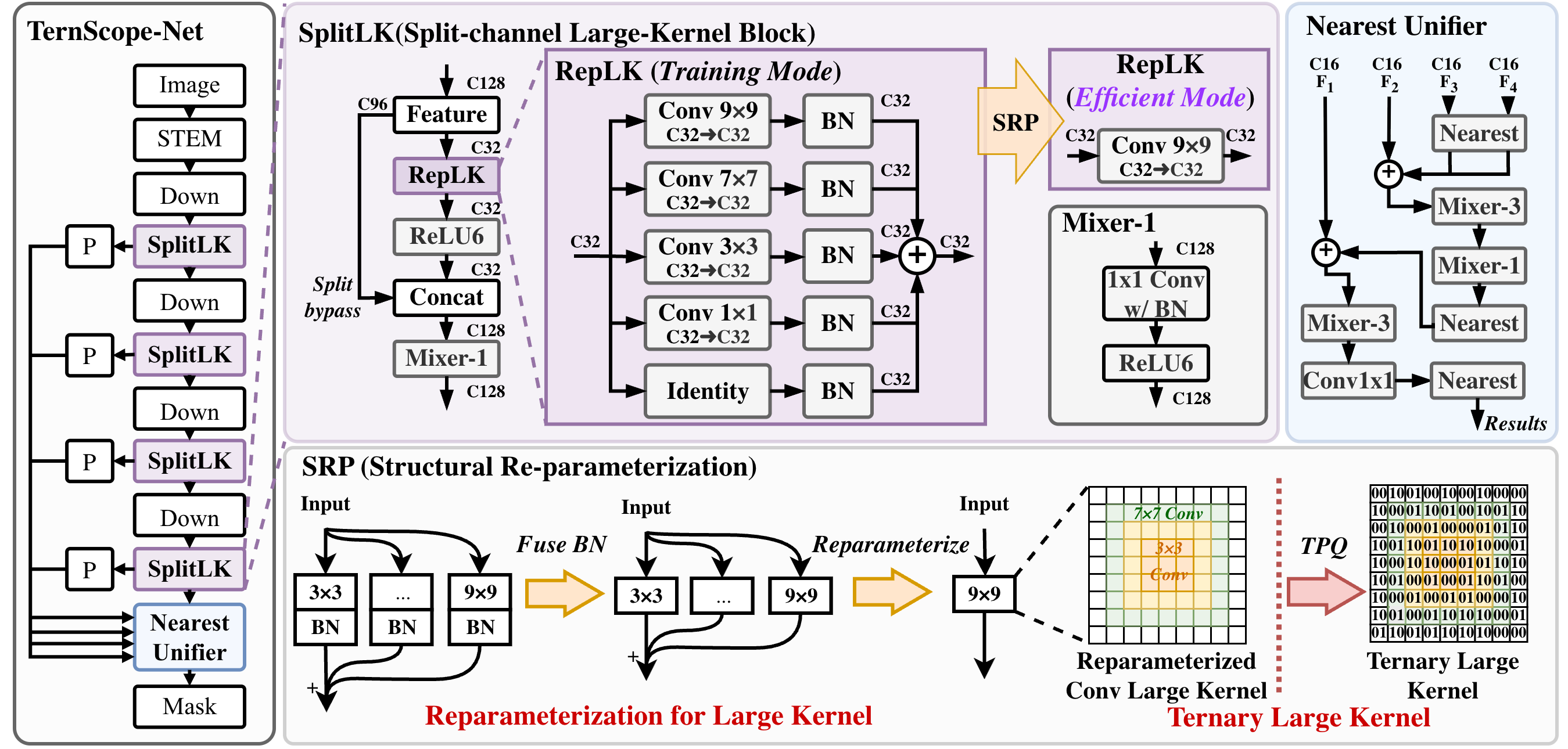}
\caption{TernScope-Net and its main components. SplitLK combines a split-channel large-kernel path with a bypass and pointwise mixing. RepLK training branches are fused into a single deployment convolution. The Nearest Unifier combines projected features from four encoder scales.}
\label{fig:network}
\end{figure*}

\ifqualafterarchitecture
\begin{figure*}[t]
\centering
\includegraphics[width=\textwidth]{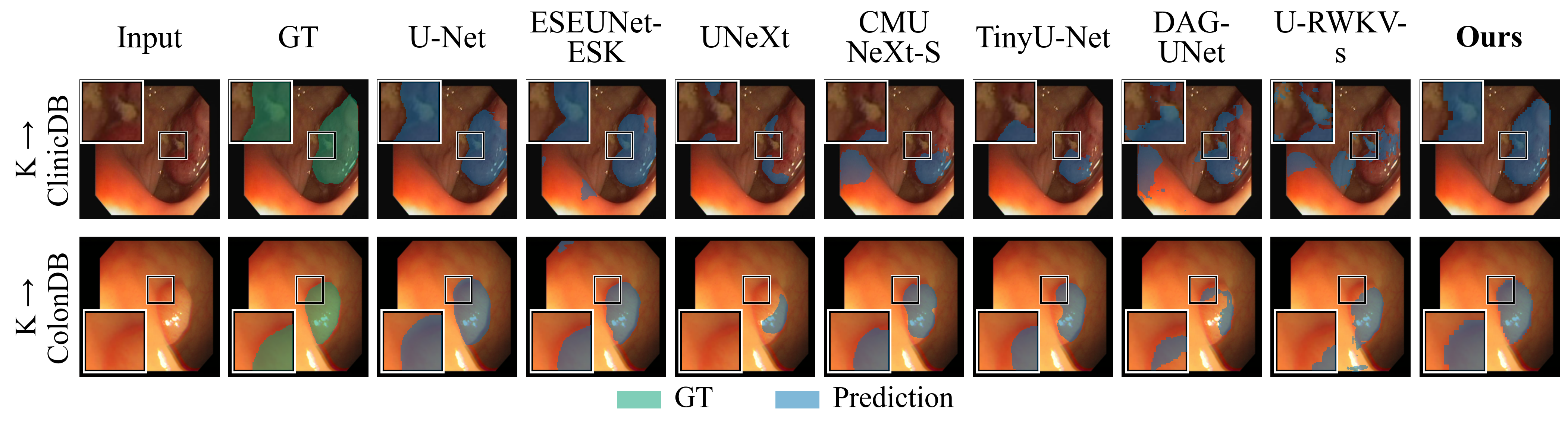}
\caption{Qualitative transfer from Kvasir-SEG to ClinicDB (top) and ColonDB (bottom), without target-domain fine-tuning. Green and blue overlays indicate ground truth and predictions, respectively. Boxed regions are enlarged in the insets.}
\label{fig:qualitative}
\end{figure*}

\fi

A compact network does not automatically translate into fast inference or a small accelerator. Execution also depends on memory access, operator organization, and the target architecture~\cite{chen2023fasternet,genc2021gemmini,chen2016eyeriss}. Structural re-parameterization helps by merging expressive training branches into efficient inference operators~\cite{ding2021repvgg,wu2026edgesegstar}. Numerical precision provides a complementary design dimension: low-bit operands reduce tensor storage and simplify arithmetic~\cite{li2016twn,zhu2017ttq,jacob2018quantization}. Combining these choices makes full-network ternary weight quantization an opportunity to address the trilemma at both model and circuit levels, provided segmentation accuracy can be retained.

The central challenge is to make a compact segmentation graph learn effectively under ternary constraints. Learned quantization scales and incremental conversion provide complementary ways to adapt networks to reduced precision~\cite{esser2020lsq,zhou2017inq}. Branch fusion changes the effective kernels, so quantizing separate training branches need not yield the same operator as quantizing their sum. Leaving the stem or prediction head at higher weight precision also retains heterogeneous convolution requirements. We instead adapt the fused deployment graph through Ternary Progressive QAT (TPQ), extending ternary weights to every convolution while retaining 8-bit activations.

TernScope realizes this model-led co-design in three connected steps (Fig.~\ref{fig:qat}). SplitLK allocates large-kernel processing to a channel subset, and a narrow decoder fuses multi-scale features. Branch fusion removes redundant execution paths before TPQ learns the final low-bit operators. A corresponding ternary array then translates the model's numerical representation into simpler computation.

We present TernScope, a fully ternary-weight segmentation
framework that jointly addresses model complexity, segmentation
accuracy, and hardware efficiency. With 0.2483\,M parameters
and 0.0949\,GMACs, TernScope outperforms its FP32 reference
across all three datasets. Using ternary PEs and shared
arithmetic resources, it reduces compute-array area,
synthesized standard-cell area, and estimated logic power
by 88.46\%, 70.28\%, and 62.82\%, respectively,
compared with the original Gemmini \cite{genc2021gemmini}.
\section{TernScope-Net}

\subsection{SplitLK: Spatial Context at Low Cost}
Large kernels expand spatial context~\cite{ding2022replknet}. TernScope-Net separates spatial coverage from channel width (Fig.~\ref{fig:network}). Its four encoder stages use 32/64/96/128 channels, while spatial convolutions process only 16/16/16/32 channels. The remaining channels bypass spatial processing and rejoin the transformed features before a pointwise mixer. For input width $C$ and processed width $C_s$, the dense spatial term falls from $HWC^2k^2$ to $HWC_s^2k^2$; pointwise mixing then exchanges information across all channels. For example, processing 32 of 128 channels reduces this spatial term to $1/16$ of its full-width counterpart; the pointwise cost remains. Large kernels of size 5, 7, and 9 operate at $32\times32$ resolution or below, where wider context is less expensive.

\begin{table*}[t]
\centering
\caption{Segmentation quality and cross-dataset transfer after training on Kvasir-SEG ($256\times256$).}
\label{tab:comparison}
\begingroup
\footnotesize
\setlength{\tabcolsep}{3.2pt}
\renewcommand{\arraystretch}{1.18}
\begin{tabular*}{\textwidth}{@{\extracolsep{\fill}}lcrrc*{6}{r}@{}}
\toprule
& & \multicolumn{2}{c}{Complexity} &
& \multicolumn{2}{c}{Kvasir-SEG}
& \multicolumn{2}{c}{$K\!\rightarrow\!$ ClinicDB}
& \multicolumn{2}{c}{$K\!\rightarrow\!$ ColonDB} \\
\cmidrule(lr){3-4}
\cmidrule(lr){6-7}
\cmidrule(lr){8-9}
\cmidrule(lr){10-11}
Model & Year & Params (M)$\downarrow$ & GMACs$\downarrow$
& W/A & Dice$\uparrow$ & IoU$\uparrow$
& Dice$\uparrow$ & IoU$\uparrow$
& Dice$\uparrow$ & IoU$\uparrow$ \\
\midrule
U-Net~\cite{ronneberger2015unet}
& 2015 & 31.0317 & 48.1670 & 32/32
& \textbf{0.8208} & \textbf{0.7303}
& \underline{0.6791} & \underline{0.5849}
& \underline{0.5360} & \underline{0.4471} \\
UNeXt~\cite{valanarasu2022unext}
& 2022 & 1.4719 & 0.5620 & 32/32
& 0.7385 & 0.6435 & 0.5207 & 0.4274 & 0.3785 & 0.2999 \\
Mobile-PolypNet~\cite{karmakar2022mobilepolypnet}
& 2022 & 0.2239 & 1.1380 & 32/32
& 0.5974 & 0.4720 & 0.4390 & 0.3269 & 0.2711 & 0.1887 \\
EGE-UNet~\cite{ruan2023egeunet}
& 2023 & 0.0534 & 0.0720 & 32/32
& 0.5882 & 0.4611 & 0.4464 & 0.3276 & 0.2804 & 0.1950 \\
CMUNeXt-S~\cite{tang2024cmunext}
& 2024 & 0.4175 & 1.0120 & 32/32
& 0.7120 & 0.6076 & 0.5029 & 0.3954 & 0.3148 & 0.2436 \\
TinyU-Net~\cite{chen2024tinyunet}
& 2024 & 0.4811 & 1.5610 & 32/32
& 0.7329 & 0.6286 & 0.5201 & 0.4329 & 0.2768 & 0.2194 \\
U-RWKV-s~\cite{ye2025urwkv}
& 2025 & 0.4704 & 0.9740$^{\dagger}$ & 32/32
& 0.6727 & 0.5608 & 0.4895 & 0.3795 & 0.2752 & 0.2035 \\
DAG-UNet~\cite{he2025dagunet}
& 2025 & 0.7358 & 0.7760 & 32/32
& 0.6785 & 0.5565 & 0.4908 & 0.3781 & 0.2484 & 0.1778 \\
ESEUNet-ESK~\cite{yu2026eseunet}
& 2026 & 0.1789 & 3.5190 & 32/32
& 0.7693 & 0.6775 & 0.5776 & 0.4759 & 0.4113 & 0.3333 \\
\midrule
\textbf{TernScope-Net (Ours)}
& --- & 0.2483 & 0.0949 & \textbf{2/8}
& \underline{0.8158} & \underline{0.7247}
& \textbf{0.7291} & \textbf{0.6290}
& \textbf{0.5945} & \textbf{0.5015} \\
\bottomrule
\end{tabular*}
\par\vspace{3pt}
\begin{minipage}{\textwidth}
\scriptsize
$K$: Kvasir-SEG, the only training dataset.
$K\rightarrow$ target: external evaluation without fine-tuning.
Test sizes: 200 / 612 / 380.
Scores are per-image means at threshold 0.5;
bold/underline denote best/second-best.
Ours uses progressive W2A8 QAT without distillation;
its parameter count excludes quantization scales.
MACs count convolution, linear layers, and matrix multiplication
at batch 1, excluding normalization, activation, and interpolation.
$^{\dagger}$WKV operators are excluded; 0.974 GMACs is partial.
\end{minipage}
\endgroup
\end{table*}
\begin{table}[!t]
\centering
\caption{Re-parameterization and ternary quantization ablation.}
\label{tab:ablation}
\begingroup
\footnotesize
\setlength{\tabcolsep}{1.8pt}\renewcommand{\arraystretch}{1.10}
\setlength{\arrayrulewidth}{0.3pt}
\begin{tabular*}{\columnwidth}{@{\extracolsep{\fill}}lrr!{\vrule width 0.3pt}lr@{}}
\toprule
Configuration & Params (M) & GMACs & Metric & Perf. (A / B / C)$\uparrow$ \\
\midrule
Multi-branch & 0.3233 & 0.1064 & Dice & (0.8102\,/\,0.7142\,/\,0.5532) \\
FP32 &  &  & IoU & (0.7218\,/\,0.6161\,/\,0.4630) \\
\midrule
Re-parameterized & 0.2483 & 0.0949 & Dice & (0.8102\,/\,0.7142\,/\,0.5532) \\
FP32 &  &  & IoU & (0.7218\,/\,0.6161\,/\,0.4630) \\
\midrule
Ternary & 0.2483 & 0.0949 & Dice & (0.0002\,/\,0.0004\,/\,0.0007) \\
PTQ &  &  & IoU & (0.0002\,/\,0.0004\,/\,0.0007) \\
\midrule
Direct ternary & 0.2483 & 0.0949 & Dice & (0.8075\,/\,0.7029\,/\,0.5636) \\
QAT &  &  & IoU & (0.7148\,/\,0.5998\,/\,0.4717) \\
\midrule
Ternary Progressive & 0.2483 & 0.0949 & Dice & \textbf{(0.8158\,/\,0.7291\,/\,0.5945)} \\
QAT (TPQ, Ours) &  &  & IoU & \textbf{(0.7247\,/\,0.6290\,/\,0.5015)} \\
\bottomrule
\end{tabular*}
\par\vspace{3pt}\begin{minipage}{\columnwidth}\scriptsize

A: Kvasir-SEG test (200); B: $K\rightarrow$ CVC-ClinicDB (612);
C: $K\rightarrow$ CVC-ColonDB (380).
$K$: Kvasir-SEG; training on $K$ only, no target fine-tuning.
W/A bits: FP32 (32/32), ternary (2/8).
Costs: $256\times256$ input; Params exclude quantization scales;
bold: ours.
\end{minipage}\endgroup

\end{table}

\begin{figure*}[t]
\centering
\includegraphics[width=\textwidth]{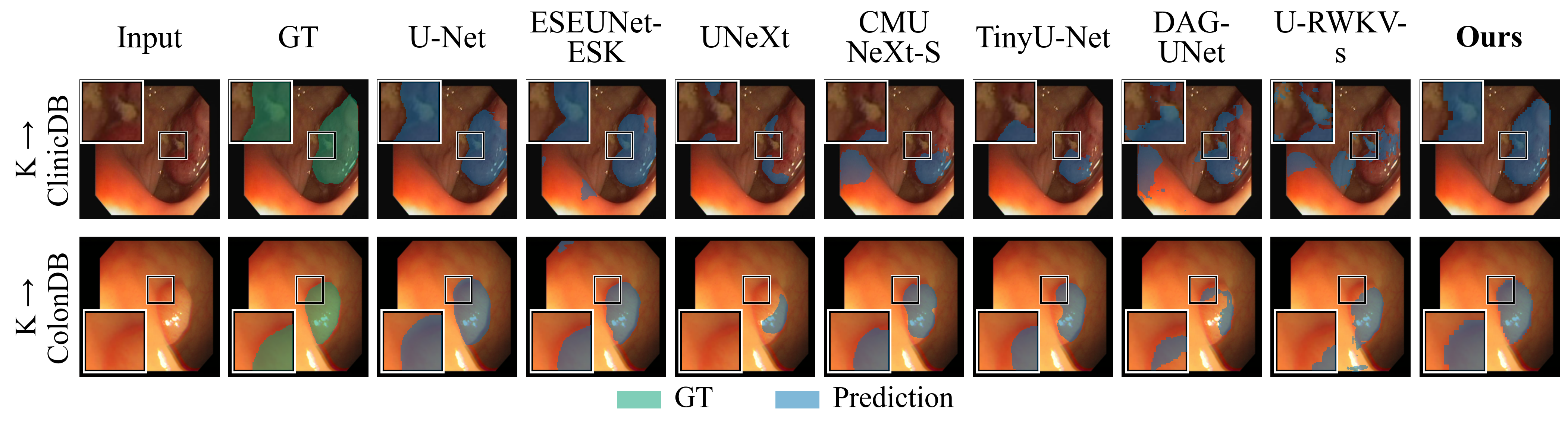}
\caption{Qualitative transfer from Kvasir-SEG to ClinicDB (top) and ColonDB (bottom), without target-domain fine-tuning. Green and blue overlays indicate ground truth and predictions, respectively. Boxed regions are enlarged in the insets.}
\label{fig:qualitative}
\end{figure*}

\subsection{Nearest Unifier: Narrow Multi-Scale Fusion}
The Nearest Unifier projects all four encoder scales to 16 channels. Nearest-neighbor resizing aligns the features for addition and convolutional refinement. The three deeper features are combined and refined before fusion with the shallow feature, so the decoder introduces semantic context before recovering fine detail. Addition keeps the fused width at 16 channels instead of growing it through concatenation. Nearest-neighbor resizing requires no learned upsampling weights. Convolutions learn spatial refinement, while dense kernels and channel widths aligned to 16 suit a common array.

\subsection{Fuse First, Then Learn Ternary Execution}
The FP32 model is first converted to its deployment topology. Batch normalization is folded into each RepLK branch, giving $(\bar W_j,\bar b_j)$, and the branches are merged as
\begin{equation}
 W_{\mathrm{rep}}=\sum_j\mathcal{P}_j(\bar W_j),\qquad
 b_{\mathrm{rep}}=\sum_j\bar b_j,
\end{equation}
where $\mathcal{P}_j$ center-pads kernels and represents the identity by a centered identity kernel. Nonlinearities remain outside the merged branches. Fusion preserves the linear operation and removes intermediate branch execution. Since quantization is nonlinear, TPQ adapts the merged kernels rather than independently quantizing branches that disappear at deployment.

For output channel $c$, the ternary weight quantizer is
\begin{equation}
 \widehat W_c=s_c\,\operatorname{clip}\!\left(\operatorname{round}(W_c/s_c),-1,1\right),
 \quad s_c>0.
\end{equation}
FP master weights and channel scales use fake
quantization with straight-through gradients. Activations
remain A8, while bias, scaling, and accumulation use wider
precision. Feature additions preserve rounding and saturation
boundaries.

TPQ follows three phases (Fig.~\ref{fig:qat}). The first 5\%
adapts the fused model in W8A8. Over the next 35\%, weights
transition to ternary from deep to shallow encoder stages,
then skip projections, decoder, input convolution, and
segmentation head. All 25 convolutions are ternary by 40\%,
with the remaining 60\% training the W2A8 graph.
Transitions use scale recalibration and brief
reduced-learning-rate adaptation. Activation scales are frozen
after final adaptation, and validation Dice selects checkpoints
only from the fully W2A8 phase.

\section{Model-to-Hardware Mapping}



\begin{figure}[t]
\centering
\includegraphics[width=\columnwidth]{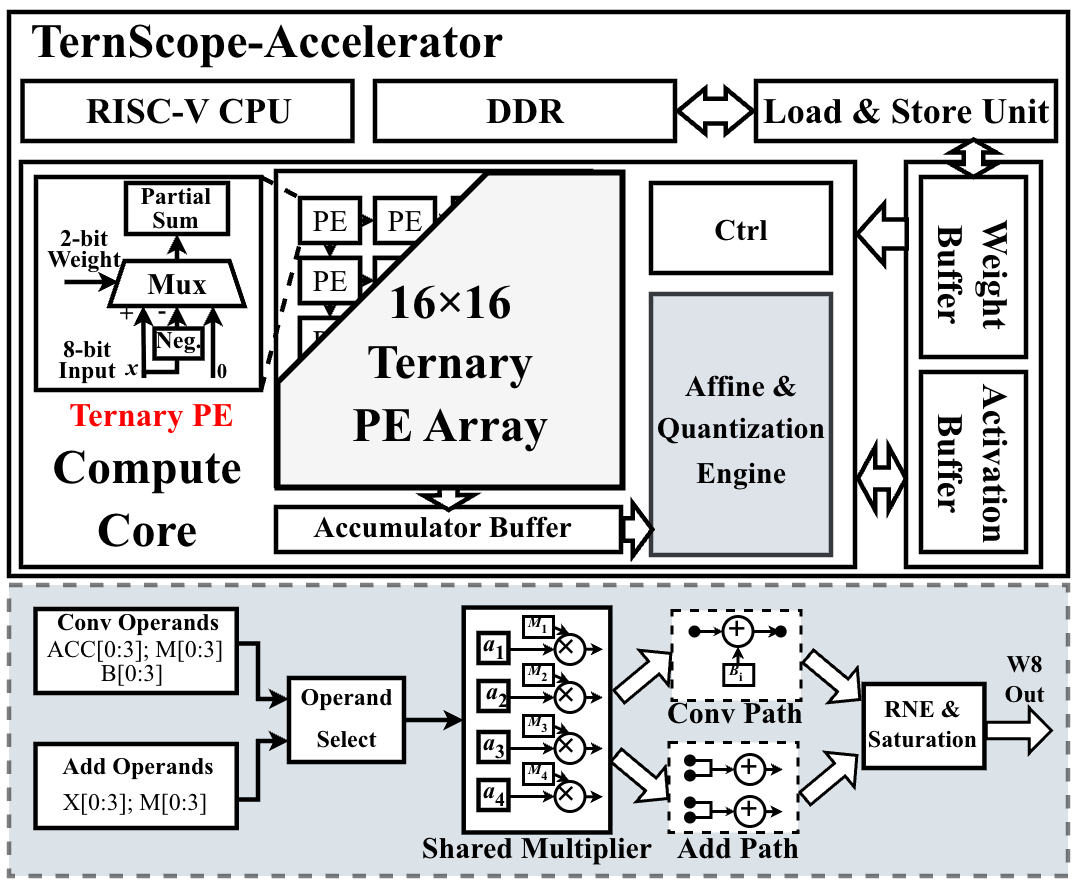}
\caption{Hardware mapping of the W2A8 deployment graph. A ternary array executes sign selection and accumulation, while the shared affine and quantization path supports convolution outputs and multi-scale feature addition.}
\label{fig:accelerator}
\end{figure}

TernScope adopts a Gemmini-based architecture comprising a
$16\times16$ ternary PE array, an accumulator buffer, and a
shared Affine \& Quantization Engine, as shown in Fig.~\ref{fig:accelerator}. Weights and activations are
loaded into on-chip buffers. The PE array performs convolution,
while the shared engine handles post-processing and
multi-scale feature addition.

The model exposes two forms of efficiency: fewer operations
after fusion and simpler arithmetic after ternarization.
Each ternary weight $t\in\{-1,0,+1\}$ selects $-a$, zero,
or $+a$ for accumulation, eliminating general multipliers
from the PE product path. Wider partial sums preserve the convolution range, followed by per-channel scaling
and bias addition.

Packed W2 weights use $1/4$ of W8 payload, while A8
keeps feature storage and transfers narrow. The Affine \&
Quantization Engine performs output post-processing. Operand selection feeds shared
multipliers, followed by bias addition for convolution or
summation of scaled features for two-source Add.
Round-to-nearest-even and saturation produce 8-bit outputs,
preserving quantization boundaries. Sparse accelerators
exploit zero operands to reduce computation and
transfers~\cite{parashar2017scnn}, motivating further
use of ternary weight sparsity.
\section{Experimental Results}
\subsection{Evaluation Protocol}
Training uses Kvasir-SEG~\cite{jha2020kvasir} only, split into 720/80/200 training/validation/test images. ClinicDB (612)~\cite{bernal2015clinicdb} and ColonDB (380)~\cite{bernal2012colondb} evaluate transfer without fine-tuning. RGB inputs are $256\times256$, normalized to $[-1,1]$; augmentation includes flips, right-angle rotations, color adjustment, and Gaussian blur. Images and masks use bilinear and nearest-neighbor resizing, respectively. Baselines use Adam, batch 16, learning rate $10^{-4}$, and binary cross-entropy plus soft Dice loss for 200 epochs. TPQ uses FP32 initialization, AdamW, batch 2, a base learning rate of $2\times10^{-4}$, and 200 epochs without distillation. Validation Dice selects checkpoints. Metrics average per-image scores at threshold 0.5. MACs count convolution, linear layers, and matrix multiplication at batch 1; Table~\ref{tab:comparison} notes the partial WKV count.

\subsection{Segmentation Quality and Quantization Ablation}
Table~\ref{tab:comparison} shows a favorable accuracy--cost tradeoff. TernScope uses approximately $1/125$ of U-Net's parameters and $1/508$ of its MACs, with in-domain Dice within 0.51 percentage points and external Dice higher by 5.00 and 5.85 points. Smaller parameter count alone does not explain this result: EGE-UNet and Mobile-PolypNet have fewer parameters, yet lower overlap scores. TernScope retains spatial context within a compact W2A8 graph, with the strongest cross-dataset Dice and IoU among the evaluated models. Fig.~\ref{fig:qualitative} provides matched examples and enlarged boundary regions.

Table~\ref{tab:ablation} separates structural simplification from low-bit adaptation. Fusion removes 23.21\% of parameters and 10.78\% of MACs while preserving FP32 overlap scores. The resulting model does not tolerate ternary PTQ: segmentation collapses despite unchanged topology. QAT restores useful predictions, and TPQ improves external Dice over direct QAT by 2.62 and 3.09 points. It also exceeds the FP32 reference on all three datasets. Low-bit adaptation is therefore essential to exploiting the fused topology.




\subsection{From Low-Bit Features to Implementation Cost}

Table~\ref{tab:hardware} compares TernScope with
Gemmini~\cite{genc2021gemmini}, a systolic-array accelerator
configured with INT8 weights and activations.
The comparison uses the same 25-convolution workload
and a $16\times16$ PE array for each design.
Both designs are synthesized using Yosys/ABC with the
Nangate45 45-nm standard-cell library.
Compared with Gemmini, TernScope reduces compute-array area
and synthesized standard-cell area by 88.457\% and 70.275\%,
respectively. These reductions stem from ternary arithmetic,
reduced accumulator width, and simplified control logic.
\begin{inlinetable}
\centering
\caption{Implementation cost for the same 25-convolution graph.}
\label{tab:hardware}
\begingroup\footnotesize
\setlength{\tabcolsep}{2.5pt}\renewcommand{\arraystretch}{1.13}
\begin{tabular*}{\columnwidth}{@{\extracolsep{\fill}}lrr@{}}
\toprule
Metric & Original Gemmini & TernScope \\
\midrule
Weight/Activation bitwidth (bits) & 8/8 & 2/8 \\
PE array & $16\times16$ & $16\times16$ \\
Library & Nangate45 & Nangate45 \\
Standard-cell area (mm$^2$) & 1.816 & 0.540 \\
Compute-array area (mm$^2$) & 0.685 & 0.079 \\
SRAM (KiB) & 320 & 344 \\
Frequency (MHz) & 100 & 100 \\
FPS & 11.902 & 102.701 \\
Logic Energy Cost (mJ/image) & 14.019 & 0.604 \\
\bottomrule
\end{tabular*}
\par\vspace{3pt}\begin{minipage}{\columnwidth}\scriptsize
Both designs use TT, 1.1\,V, $25\,^{\circ}$C and a matched Yosys/ABC flow.
Logic energy is estimated from OpenSTA vectorless power analysis.
\end{minipage}\endgroup
\end{inlinetable}

At 100\,MHz, TernScope delivers 102.701\,FPS for the evaluated
workload, compared with 11.902\,FPS for Gemmini, an
8.629$\times$ throughput improvement. Ternary weights reduce
the packed weight footprint to one quarter of the INT8
representation, shrinking model storage and easing
weight-transfer pressure. The ternary array also replaces
general multiplications with sign selection and accumulation,
reducing the logic cost of convolution.

The shared Affine \& Quantization Engine further contributes
to throughput by performing per-channel scaling, bias addition,
multi-scale feature addition, and requantization on the
accelerator. This reduces the CPU post-processing overhead
present in the evaluated Gemmini mapping. For this workload,
logic energy cost decreases from 14.019 to
0.604\,mJ per image, a 95.690\% reduction.
This co-design translates ternary model compression into
high throughput and low logic energy cost.

\section{Conclusion}
TernScope couples SplitLK, branch fusion, and TPQ to retain segmentation quality in a compact W2A8 graph. Cross-dataset evaluation and matched synthesis connect low-bit learning to a smaller compute array for edge endoscopy. A promising next step is to exploit zero-valued ternary weights through sparsity-aware execution and compressed storage, targeting lower computation and data-movement costs.

\clearpage
\bibliographystyle{IEEEtran}
\bibliography{references}
\end{document}